%% file: main.tex
\documentclass{llncs}
\let\oldbibliography\thebibliography
\renewcommand{\thebibliography}[1]{%
  \oldbibliography{#1}%
  \setlength{\itemsep}{-1.15pt}%
}
\usepackage{amssymb,amsmath}
\usepackage{array}
\usepackage{pgfplots}
\usepackage{tikz}
\usetikzlibrary{positioning,shapes,decorations.markings,arrows,backgrounds,patterns}
\usepackage{marginnote}
\usepackage{algorithm}
\usepackage{algpseudocode}
\usepackage{ifthen}
\usepackage{todonotes}
\usepackage[hidelinks]{hyperref}
\usepackage{graphicx}
\usepackage{subcaption}
\usepackage{wrapfig}
\usepackage{booktabs}
\usepackage{tabularx}
\usepackage{epsdice}

\iffalse % color
	\graphicspath{{img-color/}}
\else
	\selectcolormodel{gray}
	\graphicspath{{img-gray/}}
\fi

\title{Accelerated Model Checking of Parametric Markov Chains}

\author{Paul Gainer \and Ernst Moritz Hahn \and Sven Schewe}
\institute{University of Liverpool, UK\\
	\email{\{p.gainer,e.m.hahn,sven.schewe\}@liverpool.ac.uk}
}

\begin{document}

\maketitle

\begin{abstract}
Parametric Markov chains occur quite naturally in various applications:
they can be used for a conservative analysis of probabilistic systems (no matter how the parameter is chosen, the system works to specification);
they can be used to find optimal settings for a parameter;
they can be used to visualise the influence of system parameters; and
they can be used to make it easy to adjust the analysis for the case that parameters change.
Unfortunately, these advancements come at a cost:
parametric model checking is---or rather was---often slow.
To make the analysis of parametric Markov models scale, we need three ingredients:
clever algorithms, the right data structure, and good engineering.
Clever algorithms are often the main (or sole) selling point; and
we face the trouble that this paper focuses on -- the latter ingredients to efficient model checking.
Consequently, our easiest claim to fame is in the speed-up we have often realised
when comparing to the state of the art.
%around 1.5 to 2 orders of magnitude.
\end{abstract}

\input{definitions}

\input{introduction}
\input{preliminaries}

\input{dags}
\input{averages}

\input{experiments}

\input{conclusion}
\input{acknowledgements}

\bibliographystyle{splncs03}
\vspace{-4mm}
\bibliography{literature}

\end{document}

%% file: definitions.tex
\newcommand{\choosestate}{\mathsf{choose}}

\newcommand{\ot}{{\tiny --T--}}
\newcommand{\om}{{\tiny --M--}}

\newcommand{\enumnew}{\textsf{NumNew}}
\newcommand{\eminprod}{\textsf{MinProd}}
\newcommand{\ebfstarg}{\textsf{TargetBFS}}
\newcommand{\erandom}{\textsf{Random}}
\newcommand{\ebfs}{\textsf{BFS}}
\newcommand{\ereversebfs}{\textsf{ReverseBFS}}

\renewcommand{\pmatrix}{\mathbf{P}}
\newcommand{\states}{\mathcal{S}}
\newcommand{\reachstates}{\mathcal{B}}
\newcommand{\state}{s}
\newcommand{\estate}{\state_e}
\newcommand{\init}{{\overline{s}}}
\newcommand{\vars}{V}
\newcommand{\probzero}{\textsc{Prob}_0}
\newcommand{\ordering}{\prec}
\newcommand{\pmc}{\mathcal{D}}
\newcommand{\pre}{\mathsf{pre}}
\newcommand{\post}{\mathsf{post}}
\newcommand{\eliminate}{\textsc{Eliminate}}
\newcommand{\vol}{\mathsf{Vol}}
\newcommand{\nat}{\mathbb{N}}
\newcommand{\reals}{\mathbb{R}}
\newcommand{\evaluation}{u}
\newcommand{\poly}{g}
\newcommand{\recchar}{R}
\newcommand{\recon}{\mathcal{D}^{R}}
\newcommand{\rstates}{\states^\recchar}
\newcommand{\rinit}{\init^\recchar}
\newcommand{\rpmatrix}{\pmatrix^\recchar}
\newcommand{\cons}{\mathsf{Con}}
\newcommand{\recs}{\mathsf{Rec}}
\newcommand{\ints}{\mathsf{Int}}
\newcommand{\powerset}{\mathbb{P}}
\newcommand{\rationalfunctions}{\mathcal{F}_\vars}
\newcommand{\map}[1]{m_{#1}}
\newcommand{\domain}{\mathsf{Dom}}
\newcommand{\elim}{\mathsf{Eliminate}}
\newcommand{\infected}{\mathsf{Infected}}
\newcommand{\stateelim}{\mathsf{StateElimination}}
\newcommand{\elimset}{\mathsf{Elim}}
\newcommand{\neigh}{\mathsf{Neigh}}
\newcommand{\reach}[1]{\mathsf{reach}_{#1}}
\newcommand{\graph}{\mathcal{G}}
\newcommand{\zeroconf}{Z}
\newcommand{\zinit}{\mathrm{i}}
\newcommand{\zok}{\mathrm{ok}}
\newcommand{\zerr}{\mathrm{err}}
\newcommand{\rvariable}{X}
\newcommand{\rfunc}{r}
\newcommand{\expect}{E}
\newcommand{\paths}{\mathsf{Paths}}
\newcommand{\pmeasure}{\mathsf{Pr}}
\newcommand{\pexpect}{\mathsf{Exp}}
\renewcommand{\path}{\omega}
\newcommand{\stovar}{X}

\newcommand{\propreach}{\mathsf{Reach}}
\newcommand{\propacc}{\mathsf{Acc}}
\newcommand{\proplra}{\mathsf{LRA}}
\newcommand{\propreturn}{\mathsf{Return}}

%% file: introduction.tex
\section{Introduction}

The analysis of parametric Markov models is a young and growing field of research.
As not only the research direction but also the term `parametric Markov models' is attractive, it has been used for various generalisations of traditional Markov models.
We use Markov chains, where the parameter is used to determine the probabilities and rewards, such that we can reason about the likelihood of obtaining simple
temporal properties like safety and reachability as well as standard reward functions, such as long-run average.

What we do \emph{not} intend to do in this paper is to use parameters to change the size of the system or the shape of the Markov chain.
(The latter can, of course, be encoded by using parameters to assign a probability of $0$ to an edge, effectively removing it.
This would, however, come at the cost of efficiency and is not what we want to use the parameters for.)

Using parameters to describe the probabilities of transitions is not quite as easy as it sounds: even when parameters appear in a simple way, like `$p$' or `$1-p$',
the terms that represent the likelihood of obtaining a temporal property or an expected reward can quickly become quite intricate.
One ends up with rational functions.
We make a virtue of necessity by using this as a motivation to allow for using rational functions of the occurring parameters to represent the probabilities and payoffs.

To allow for an efficient analysis of such complex parametrised systems, we have taken a look at different strategies for the evaluations
of---parametric and non-parametric---Markov chains, and considered their suitability for our purposes.
We found the stepwise elimination of vertices from a model to be the most attractive approach to port.

Broadly speaking, this approach works like the transformation from finite automata to regular expressions:
a vertex is removed, and the new structure has all successors of this state as---potentially new---successors of the predecessors of  this  vertex.
In the transformation from finite automata to regular expressions, one changes the expressions on the edges, while
we adjust the probabilities and, if applicable, the rewards on the edges.

When using this approach with explicit probabilities and rewards, one ends up with a Directed Acyclic Graph (DAG) structure in the evaluation.
This DAG structure has been exploited to reduce the cost of re-calculating the probabilities for simple temporal properties or expected rewards, and
it proves that it also integrates nicely into our framework, where the probabilities and rewards are provided as rational functions.
In fact it integrates so naturally that it seems surprising in hindsight that it has not been discovered earlier.

The natural connection occurs when choosing a similar data structure to represent the rational functions that represent the probabilities and rewards.
To make full use of the DAG structure that comes with the elimination, we represent these functions in the form of arithmetic circuits---which are essentially DAGs.
We have integrated the resulting representation organically in a small extension of ePMC, and tested it on a range of case studies.
We have obtained a speed-up of a hefty factor of 20 to 120 when compared to storing functions in terms of coprime numerator and denominator polynomials.
%, using a simple efficient heuristic to simplify the occurring fractions.

\paragraph*{\bf Related work. }
For (discrete-time) Markov chains (MCs), Daws~\cite{daws2004symbolic} has devised a language-theoretic approach to solve this problem.
In this approach, the transition probabilities are considered as letters of an alphabet.
Thus, the model can be viewed as a finite automaton.
Then, based on the state elimination method~\cite{hopcroft2008introduction}, a regular expression that describes the language of such an automaton is calculated.
In a post-processing step, this regular expression is recursively evaluated, resulting in a rational function over the parameters of the model.
One of the authors has been involved in extending and tuning this method~\cite{hahn2011probabilistic} so as to operate with rational functions, which are stored as coprime numerator and denominator polynomials rather than with regular expressions.

The process of computing a function that describes properties (like reachability probabilities or long-run average rewards) that depend on model parameters is often costly.
However, once the function has been obtained, it can very efficiently be evaluated for given parameter instantiations.
Because of this, parametric model checking of Markov models has also attracted attention in the area of runtime verification,
where the acceptable time to obtain values is limited~\cite{calinescu2012self,filieri2011run}.

Other works in the area are centred around deciding the validity of boolean formulas depending on the parameter range using SMT solvers or
extending these techniques to models that involve nondeterminism~\cite{DehnertJ0CVKAB16,HahnHZ11,Cubuktepe0JKPPT17,QuatmannD0JK16}.

\begin{wrapfigure}[16]{r}{0.6\textwidth}
	\vspace{-15pt}
	\begin{center}
          \input{dice}
        \end{center}
        \caption{Simulating a biased dice by a biased coin.\label{fig:dice-model}}
\end{wrapfigure}
As an example for a parametric model, consider Figure~\ref{fig:dice-model}.
Knuth and Yao~\cite{KY76} have shown how a six-sided dice can be simulated by repeatedly tossing a coin.
The idea is to build a Markov chain with transition probabilities of only $0.5$ or $1$.
Borrowing a model from the PRISM website, we have extended this example to a biased dice, simulated by tossing a biased coin.
With probability $x$ we see heads, while with probability $1-x$ we see tails.
This way, we move around in the Markov chain until we obtain a result.
        
\paragraph*{\bf Organisation of the paper. }
After formalising our setting in Section \ref{sec:preliminaries},
we describe how we exploit DAGs in the representation of rational functions, and exploit them using synergies with the DAG-style state elimination technique, in Section \ref{sec:dag}.
We then describe how to expand this technique to determine long-run average rewards in Section \ref{sec:lra}.
In Section \ref{sec:experiments}, we evaluate our approach on a range of benchmarks, and discuss the results briefly in Section \ref{sec:conclude}. 

%% file: dice.tex
\begin{tikzpicture}
  \tikzstyle{regular} = [draw, shape = circle, inner sep = 0pt, minimum size = 0.7cm];
  \tikzstyle{result} = [inner sep = 0pt];
  \tikzstyle{arrow} = [-stealth, thick];
  \tikzstyle{abovelabel} = [pos = 0.5, above];
  \tikzstyle{belowlabel} = [pos = 0.35, below];

  \node (init) at (-0.7,0) {};
  
  \node[regular] (s0) at (0,0) {$0$};
  \node[regular] (s1) at (2,1) {$1$};
  \node[regular] (s2) at (2,-1) {$2$};
  
  \node[regular] (s3) at (4,2.25) {$3$};
  \node[regular] (s4) at (4,0.75) {$4$};
  \node[regular] (s5) at (4,-0.75) {$5$};
  \node[regular] (s6) at (4,-2.25) {$6$};

  \node[result] (d1) at (6,2.5) {\Huge\epsdice{1}};
  \node[result] (d2) at (6,1.5) {\Huge\epsdice{2}};
  \node[result] (d3) at (6,0.5) {\Huge\epsdice{3}};
  \node[result] (d4) at (6,-0.5) {\Huge\epsdice{4}};
  \node[result] (d5) at (6,-1.5) {\Huge\epsdice{5}};
  \node[result] (d6) at (6,-2.5) {\Huge\epsdice{6}};

  \draw[->] (init) -- (s0);
  \draw[->] (s0) -- node[above] {$x$} (s1);
  \draw[->] (s0) -- node[below] {$1-x$} (s2);
  \draw[->] (s1) -- node[below] {$x$} (s3);
  \draw[->] (s1) -- node[below] {$1-x$} (s4);
  \draw[->] (s2) -- node[above] {$x$} (s5);
  \draw[->] (s2) -- node[above] {\ \ \ $1-x$} (s6);
  \draw[->] (s3) edge[bend right=40] node[above] {$x$} (s1);
  \draw[->] (s3) -- node[above] {$1-x$} (d1);
  \draw[->] (s4) -- node[above] {$x$} (d2);
  \draw[->] (s4) -- node[below] {$1-x$} (d3);
  \draw[->] (s5) -- node[above] {$x$} (d4);
  \draw[->] (s5) -- node[below] {$1-x$} (d5);
  \draw[->] (s6) -- node[below] {$x$} (d6);
  \draw[->] (s6) edge[bend left=40] node[below] {$1-x$} (s2);
\end{tikzpicture}

%% file: preliminaries.tex
\section{Preliminaries}
\label{sec:preliminaries}

\subsection{Parametric Markov chains with state rewards}
%We first describe the state elimination method of
%Hahn~\cite{hahn2011probabilistic} for parametric
%Markov Chains (PMCs), and then
%introduce an algorithm that substantially reduces the
%cost of recomputation of the parametric
%reachability probability for a reconfigured PMC. First we
%give some general definitions.
%Given a function $f$ we denote the domain of $f$ by
%$\domain(f)$. We use the notation
%$f \oplus f^\prime = f \restriction_{\domain(f)
%		\setminus \domain(f^\prime)} \cup f^\prime$
%to denote the overriding union of $f$ and $f^\prime$.

Let $\vars = \{v_1, \ldots, v_n\}$ denote a set of variables
over~$\reals$.
%An evaluation for $\vars$ is a partial
%function $\evaluation \colon \vars \to \reals$.
A \emph{polynomial} $\poly$ over $\vars$ is a sum of monomials
\begin{align*}
	\poly(v_1, \ldots, v_n) = \sum_{i_1, \ldots, i_n}
			a_{i_1},\ldots ,_{i_n} v_1^{i_1} \ldots v_n^{i_n},
\end{align*}
where each $i_j \in \nat$ and each $a_{i_1},\ldots,_{i_n} \in \reals$.
A \emph{rational function} $f$ over a set of variables $\vars$
is a fraction
$f(v_1,\ldots,v_n) = \frac{f_1(v_1,\ldots,v_n)}{f_2(v_1,\ldots,v_n)}$
of two polynomials $f_1, f_2$ over $\vars$. We denote the set
of rational functions from $\vars$ to $\reals$ by $\rationalfunctions$.

\begin{definition}
A \emph{parametric Markov chain} (PMC) is a tuple
$\pmc = (\states, \init, \pmatrix, \vars)$, where $\states$
is a finite set of states, $\init$ is the initial state,
$\vars = \{v_1, \ldots, v_n\}$ is a finite set of parameters,
and $\pmatrix$ is the probability matrix
$\pmatrix \colon \states \times \states \to \rationalfunctions$.
A \emph{path} $\path$ of a PMC $\pmc = (\states, \init, \pmatrix, \vars)$
is a non-empty finite, or infinite, sequence $\state_0, \state_1, \state_2, \ldots$
where $\state_i \in \states$ and $\pmatrix(\state_i, \state_{i+1}) > 0$ for $i \geqslant 0$.
We let $\Omega$ denote the set of infinite paths.
With $\pmeasure_\state$, we denote the parametric probability measure over $\Omega$ assuming that we start in state $\state$, with $\pmeasure = \pmeasure_\init$.
We use $\pexpect_\state$, $\pexpect$ to denote according expectations.
With
$\stovar(\pmc)_{\state,i}\colon \Omega \to \states$,
$\stovar(\pmc)_{\state,i}(\state_0, \state_1, \ldots) = \state_i$
we denote the random variable expressing the state occupied at step $i\geq 0$, and let
$\stovar(\pmc)_{i} = \stovar(\pmc)_{\init,i}$.
\end{definition}

\begin{definition}
Given a PMC $\pmc = (\states, \init, \pmatrix, \vars)$,
the \emph{underlying graph} of $\pmc$
is given by $\graph_\pmc = (\states, E)$ where 
$E = \{(\state, \state^\prime) \mid
\pmatrix(\state, \state^\prime) > 0 \}$.
A \emph{bottom strongly connected component (BSCC)} is a set $A \subseteq \states$ such that in the underlying graph each state $\state_1 \in A$ can reach each state $\state_2 \in A$ and there is no $\state_3 \in \states \setminus A$ reachable from $\state_1$.
\end{definition}
Given a state $\state$, we denote the set
of all immediate predecessors and successors of $\state$
in the underlying graph of $\pmc$ by $\pre_\pmc(\state)$ and
$\post_\pmc(\state)$, respectively, excluding $\state$ itself.
%, and we define the
%\emph{neighbourhood} of $\state$ as
%$\neigh(\state) = \state \cup \pre_\pmc(\state) \cup \post_\pmc(\state)$.
We write $\reach{D}(\state, \state^\prime)$ if
$\state^\prime$ is reachable from $\state$ in the underlying graph
of $D$. 

\begin{algorithm}[tb]
	\caption{Parametric Reachability Probability for PMCs}
	\label{alg:stateelim}
	\begin{algorithmic}[1]
		\Procedure{\textsc{StateElimination}}{$\pmc, \reachstates$}
			\State{\textbf{requires:}}
			A PMC $\pmc = (\states, \init, \pmatrix, \vars)$ and set of target
			states $\reachstates \subseteq \states$, where $\reach{D}(\init, \state)$
                        holds for all $\state \in \states$.
			\State $E \gets \states $
			\While{$E \ne \emptyset$}			
				\State $\estate \gets \choosestate(E)$
				\State $E \gets E \setminus \{\estate\}$
                                \ForAll{$\state \in \post_\pmc(\estate)$}
                                \State $\pmatrix(\estate,\state) \gets \pmatrix(\estate,\state) / (1-\pmatrix(\estate,\estate))$
                                \EndFor
                                \State $\pmatrix(\estate,\estate) \gets 0$
				\ForAll{$(\state_1, \state_2) \in \pre_\pmc(\estate)
						\times \post_\pmc(\estate)$}
					\State $\pmatrix(\state_1, \state_2) \gets
							\pmatrix(\state_1, \state_2) + \pmatrix(\state_1, \estate)
							\pmatrix(\estate, \state_2)$		
				\EndFor
                                \If{$\estate \neq \init \wedge \estate \notin \reachstates \wedge \post_\pmc(\estate) \neq \emptyset$}
				\State $\elim(\pmc, \estate)$
				  \textit{// remove $\estate$ and incident transitions from $\pmc$}
                                \EndIf
			\EndWhile
			\State \textbf{return} $\sum_{s \in \reachstates }\pmatrix(\init, \state)$
		\EndProcedure
	\end{algorithmic}
\end{algorithm}

Given a PMC
$\pmc = (\states, \init, \pmatrix, \vars)$
we are interested in computing the function that represents the probability of reaching some set of target states $\reachstates \subset \states$.
\begin{align*}
\propreach(\pmc,\reachstates) = \pmeasure \left[ \exists i \geq 0. \stovar(\pmc)_{\init,i} \in \reachstates  \right]
\end{align*}
Our base algorithm to obtain this value is described in Algorithm~\ref{alg:stateelim}.
A state $\estate \in \states$ is selected, and then eliminated by considering each pair
$(\state_1, \state_2) \in \pre_\pmc(\estate) \times \post_\pmc(\estate)$
and updating the existing probability $\pmatrix(\state_1, \state_2)$ by the probability of reaching $\state_2$ from $\state_1$ via $\estate$.
Heuristics to determine the order in which states are chosen for elimination by the $\choosestate$ function are discussed in Section~\ref{sec:heuristics}.

\begin{definition}
  A \emph{parametric reward function} for a PMC
  $\pmc = (\states, \init, \pmatrix, \vars)$
  is a function $\rfunc \colon \states \to \rationalfunctions$.
\end{definition}
The reward function labels states in $\pmc$ with a rational function
over $\vars$ that corresponds to the reward that is gained if that state
is visited.
Given a PMC $\pmc = (\states, \init, \pmatrix, \vars)$ and a reward function $\rfunc \colon \states \to \rationalfunctions$, we are interested in the \emph{parametric expected accumulated reward}
defined as
\begin{align*}
\propacc(\pmc,\rfunc) = \pexpect \left[ \sum_{i=0}^\infty \rfunc(\stovar(\pmc))_{\init,i} \right] 
\end{align*}
or a variation~\cite{kwiatkowska2007stochastic}, the \emph{parametric expected accumulated reachability reward} given $\reachstates \subseteq \states$ defined as
\begin{align*}
\propacc(\pmc,\rfunc,\reachstates) = \pexpect \left[ \sum_{i=0}^{\{j \mid \stovar(\pmc)_{\init,j} \in \reachstates \}} \rfunc(\stovar(\pmc))_{\init,i} \right] .
\end{align*}
This can, however, be transformed to the former.

Algorithm~\ref{alg:stateelim} can be extended to compute the parametric expected accumulated reward. In addition to updating the probability matrix for each predecessor and successor pair, we also update the reward function as follows:
\begin{align*}
\rfunc(\state_1) \gets \rfunc(\state_1) + \pmatrix(\state_1, \estate) 
	\frac{\pmatrix(\estate, \estate)}{1 - \pmatrix(\estate, \estate)}
	\rfunc(\estate).
%\rfunc(\state_2) \gets \rfunc(\state_2) + 
%	\frac{\pmatrix(\state_1, \estate) \pmatrix(\estate, \state_2)}
%				{1 - \pmatrix(\estate, \estate)}
%	\rfunc(\estate).
\end{align*}
The updated value for $\rfunc(\state_1)$ reflects the reward that would be accumulated if a transition would be taken from $\state_1$ to $\estate$, where the expected number of self-loops would be $\frac{\pmatrix(\estate, \estate)}{1 - \pmatrix(\estate, \estate)}$.
Upon termination, the algorithm returns the value
$\rfunc(\init)$.

%% file: dags.tex
\section{Representing Formulas using Directed Acyclic Graphs}
\label{sec:dag}
In existing tools for parametric model checking of Markov models, rational functions have traditionally been represented in the form $f(v_1,\ldots,v_n) = \frac{f_1(v_1,\ldots,v_n)}{f_2(v_1,\ldots,v_n)}$, where $f_1(v_1,\ldots,v_n)$ and $f_2(v_1,\ldots,v_n)$~\cite{hahn2010param,dehnert2017storm,kwiatkowska2011prism} are coprime.
As a result, for some cases the representations of such functions are very short. Often, during the state elimination phase, large common factors can be cancelled out, such that one can operate with relatively small functions throughout the whole algorithm.
There are, however, many cases without---or with very few---large common factors.
The nominator-denominator representations then become larger and larger during the analysis. 
In this case, the analysis is slowed down severely, mostly by the time taken for the cancellation of common factors.
Cancelling out such factors is non-trivial, and indeed a research area in itself.
In addition, if formulas become large, this can also lead to out-of-memory problems.

To overcome this issue, we propose the representation of rational functions by  arithmetic circuits.
These arithmetic circuits are directed acyclic graphs (DAG).
Terminal nodes are labelled with either a variable of the set of parameters $\vars$, or with a rational number.
Non-final nodes are labelled with a function to be applied on the nodes it has edges to.
In our setting, we require two unary functions, additive inverse and multiplicative inverse, and two binary functions, addition and multiplication.
All functions used are represented using a single DAG, and a function is represented by a reference to a node of this common DAG.

This representation has two advantages.
Firstly, all operations are practically constant time:
to apply an operator on two functions, one simply introduces a new node labelled with the according operator, with edges pointing to the two nodes to connect.
In particular, we do not have to use expensive methods to cancel out common factors.
Secondly, because we are using a DAG and not a tree, common sub-expressions can be shared between different formulas, which is not possible when representing rational functions in terms of two polynomials represented as a list of monomials.

\begin{wrapfigure}[26]{r}{0.4\textwidth}
	\vspace{-36pt}
	\begin{center}
          \input{dice-dag}
        \end{center}
        \caption{Probability of rolling \epsdice{6}.\label{fig:dice-dag}}
\end{wrapfigure}
For illustration, let us consider the example from Figure~\ref{fig:dice-model}.
We analyse the probability that the final result is \epsdice{6}.
This probability can be described by the function
\begin{align*}
  \frac{-x^2+2x-1}{x-2} .
\end{align*}
In our DAG-based representation, we would represent the function as in Figure~\ref{fig:dice-dag};

When operating with arithmetic circuits, there are a number of ways to reduce their memory footprint, which will, however, lead to a higher running time.
The simplest one is that, while creating a new node to represent a function, it might turn out that there already exists a node with exactly the same operator, and exactly the same operand.
In this case, it is better to drop the newly created node and use a reference to the existing node to counter the growth of the DAG.
In case we use hash maps for the lookup, we can also still keep the overhead close to constant time.
Another optimisation is to use simple algebraic equivalences.
This includes computing the values of constant functions.
E.g.\ instead of creating a node representing $2+3$ we introduce a new terminal node labelled $5$, and
if we are about the create a new node for $y+x$ but we already have a node for $x+y$ we reuse this node instead.
We also take the additive and multiplicate neutral elements into account
(rather than creating a new node for $0+x$, we return the one for $x$, and the like).
Another optimisation method is to evaluate functions of the DAG at random points and then to identify functions if the result of this evaluation is the same.
Using the Schwartz-Zippel Lemma~\cite{Schwartz80,Zippel79,DeMillo1978APR},
we can then bound the probability that we mistakenly identify two nodes although they do not represent the same function.
We can again minimise the overheads incurred by this method by using hash maps.

Arithmetic circuits sometimes become very large, consisting of millions of nodes.
This way, they cannot serve as a concise, human-readable description of the analysis result.
Compared to performing a non-parametric analysis, it is however often still beneficial to obtain a function representation in this form.
Even for large dags, obtaining evaluating parameter instantiations is very fast, and linear in the number of DAG nodes.
This is useful in particular if a large number of points is required, for instance for plotting a graph.
In this case, results can be obtained much faster than using non-parametric model checking, as demonstrated in Section~\ref{sec:experiments}.
In particular, for any instantiation, values can be obtained in the same, predictable, time.
This is quite in contrast to value iteration, where the number of iterations required to obtain a certain precision varies with the concrete values of parameters.

For this reason, parametric model checking is particularly useful for online model checking or runtime verification~\cite{calinescu2012self}.
Here, one can precompute the DAG before running the actual system, while concrete values can be instantiated at runtime,
with a running time that can be precisely calculated offline.
Using arithmetic circuits expands the range of systems for which this method is applicable.
Evaluation of parameter instantiations can be performed using exact arithmetic or floating-point arithmetic.
From our experience, the quality of the floating-point results using DAGs is often better than the one using the representation of rational functions as coprime numerator and denominator,
which has been used so far in known implementations.
The reason is that, in the latter approach, one often runs into numerical problems such as cancellation, which often forces the use of expensive exact arithmetic to be used for evaluation.
The DAG-based method seems to be more robust against such problems.

It has recently come noted by the verification community that the usual way in which value iteration is implemented is not safe, and solutions have already been proposed~\cite{BKLPW17}.
While this solves the problem, it requires more complex algorithms and leads to increased model checking time.
In case arithmetic circuits are used, it is easy to obtain conservative upper and lower bounds for parameter instantiations.
One only has to use interval arithmetic and provide implementations for the basic operations used (addition, multiplication, additive and multiplicative inverse).
The increase in the time to evaluate functions is small.
In our experiments, the largest interval diameter we have obtained is around $10^{-13}$.

%% file: dice-dag.tex
\begin{tikzpicture}[>=latex',line join=bevel,yscale=0.4,xscale=0.6]
  \tikzstyle{inner} = [draw, shape = circle, inner sep = 0pt, minimum size = 0.7cm];
  \node (v30) at (52.197bp,617.74bp) [inner] {$\cdot$};
  \node (v23) at (129.2bp,238.75bp) [inner] {$\cdot$};
  \node (v27) at (32.197bp,459.5bp) [inner] {$/$};
  \node (v26) at (129.2bp,387.5bp) [inner] {$+$};
  \node (v25) at (129.2bp,315.5bp) [inner] {$-$};
  \node (v24) at (27.197bp,238.75bp) [inner] {$\cdot$};
  \node (v29) at (27.197bp,536.24bp) [inner] {$\cdot$};
  \node (v1) at (170.2bp,18.0bp) [draw,rectangle] {$1$};
  \node (v2) at (73.197bp,18.0bp) [draw,rectangle] {$x$};
  \node (v3) at (97.197bp,90.0bp) [inner] {$-$};
  \node (v4) at (98.197bp,162.0bp) [inner] {$+$};
  \node (start30) at (52.197bp,678.29bp) [] {};
  
  \draw [->] (v24) ..controls (53.544bp,210.27bp) and (67.571bp,195.11bp)  .. (v4);
  \draw [->] (v3) ..controls (88.706bp,64.527bp) and (85.457bp,54.781bp)  .. (v2);
  \draw [->] (v29) ..controls (12.672bp,504.53bp) and (7.5644bp,490.57bp)  .. (5.1966bp,477.5bp) .. controls (-8.0504bp,404.35bp) and (7.8917bp,317.07bp)  .. (v24);
  \draw [->] (v29) ..controls (29.215bp,505.27bp) and (29.801bp,496.27bp)  .. (v27);
  \draw [->] (v23) ..controls (111.25bp,211.05bp) and (106.53bp,199.88bp)  .. (v4);
  \draw [->] (v23) ..controls (123.06bp,206.45bp) and (118.51bp,195.15bp)  .. (v4);
  \draw [->] (v26) ..controls (129.2bp,361.63bp) and (129.2bp,352.47bp)  .. (v25);
  \draw [->] (start30) ..controls (52.197bp,672.0bp) and (52.197bp,661.52bp)  .. (v30);
  \draw [->] (v30) ..controls (65.838bp,565.24bp) and (78.197bp,508.64bp)  .. (78.197bp,459.5bp) .. controls (78.197bp,459.5bp) and (78.197bp,459.5bp)  .. (78.197bp,315.5bp) .. controls (78.197bp,271.26bp) and (86.464bp,220.46bp)  .. (v4);
  \draw [->] (v25) ..controls (129.2bp,289.74bp) and (129.2bp,280.72bp)  .. (v23);
  \draw [->] (v24) ..controls (40.319bp,175.78bp) and (58.165bp,90.133bp)  .. (v2);
  \draw [->] (v27) ..controls (62.688bp,436.86bp) and (87.889bp,418.16bp)  .. (v26);
  \draw [->] (v30) ..controls (42.863bp,587.31bp) and (39.787bp,577.29bp)  .. (v29);
  \draw [->] (v4) ..controls (97.837bp,136.13bp) and (97.71bp,126.97bp)  .. (v3);
  \draw [->] (v26) ..controls (145.06bp,361.63bp) and (152.44bp,347.25bp)  .. (156.2bp,333.5bp) .. controls (176.84bp,257.86bp) and (174.95bp,164.26bp)  .. (v1);
  \draw [->] (v4) ..controls (120.43bp,139.77bp) and (133.4bp,126.8bp)  .. (v1);
\end{tikzpicture}

%% file: averages.tex
\section{Computation of Fractional Long-Run Average Values}
\label{sec:lra}
Consider a PMC $\pmc = (\states, \init, \pmatrix, \vars)$ together with two reward functions $\rfunc_u \colon \states \to \rationalfunctions$ and $\rfunc_l \colon \states \to \rationalfunctions$.
The problem we are interested in is computing the value \emph{fractional long-run average reward}~\cite{BloemCGHHJKK14,EssenJ11}
\begin{align*}
  \proplra(\pmc,\rfunc_u,\rfunc_l,\state) {=} \pexpect\!\!\left[\lim_{n\to\infty} \frac{\sum_{i=0}^n \rfunc_u(\stovar(\pmc)_{\state,i})}{\sum_{i=0}^n \rfunc_l(\stovar(\pmc)_{\state,i})} \right]\!,
  \proplra(\pmc,\rfunc_u,\rfunc_l) {=} \proplra(\pmc,\rfunc_u,\rfunc_l,\init) .
\end{align*}
In a simple case, $\rfunc_l(\cdot) = 1$, which means that we compute the \emph{long-run average reward}
\begin{align*}
\pexpect \left[\lim_{n\to\infty} \frac{1}{n+1} \sum_{i=0}^n \rfunc_u(\stovar(\pmc)_{\state,i}) \right] ,
\end{align*}
where each step is assumed to take the same amount of time.
Solution methods for this property has been implemented (but to the best of our confidence, not been published) for parametric models in PRISM and Storm.
The fractional long-run average reward is more general and allows to express values like the average energy usage per task performed more easily.
Given a reward structure $\rfunc \colon \states \to \rationalfunctions$, we define the \emph{recurrence reward} as
\begin{align*}
\propreturn(\pmc,\rfunc,\state)  = \pexpect \left[ \sum_{i=0}^{\min \{ j \mid \stovar(\pmc)_{\state,j} = s \wedge j > 0 \}} \!\!\!\!\!\!\!\!\!\!\!\!\!\!\!\!\!\!\!\! \rfunc(\stovar(\pmc)_{\state,i}) \right],
\propreturn(\pmc,\rfunc)  = \propreturn(\pmc,\rfunc,\init) .
\end{align*}
It is known~\cite{Cox67}
%\url{http://www.columbia.edu/~ks20/stochastic-I/stochastic-I-RP.pdf})
that this value is the same for all states of a BSCC.
Furthermore, for $\rfunc_l(\cdot) = 1$ we have
\begin{align*}
\frac{\propreturn(\pmc,\rfunc_u,\state)}{\propreturn(\pmc,\rfunc_l,\state)} = \proplra(\pmc,\rfunc_u,\rfunc_l,\state) ,
\end{align*}
which immediately extends to the general case.

In Section~\ref{sec:preliminaries}, we have discussed how state elimination can be used to obtain values for the expected accumulated reward values.
For this, we have repeatedly eliminated states so as to bring the PMC of interest into a form in which reward values can be obtained in a trivial way.
It is easy to see that the transformations for the expected accumulated rewards also maintains the recurrence rewards.
After having handled each state of our model, we have two possible outcomes.
\begin{wrapfigure}[11]{r}{0.6\textwidth}
	\vspace{-34pt}
	\begin{center}
	\begin{tikzpicture}
		\tikzstyle{nodestyle} = [draw, shape = circle, 	inner sep = 0pt, minimum size = 0.7cm];
		\tikzstyle{arrow} = [-stealth, thick];
		\tikzstyle{abovelabel} = [pos = 0.5, above];
		\tikzstyle{belowlabel} = [pos = 0.35, below];
		\def \spacing {0.78cm}
		\def \bendangle {50}
		\def \labelshift {(0.6, 0.1)}

		\def \labelshift {(0.0, 0.0)}
		\node (1) [nodestyle] {};
		\node (foo) [left = 0.5cm of 1] {};
		\draw [arrow] (foo)--(1);
		\node (2) [nodestyle, shift={(0, -2)}] {};
		\node (bar) [left = 0.5cm of 2] {};
		\draw [arrow] (bar)--(2);
		\node (3) [nodestyle, shift={(1.6, -1)}] {};
		\node (4) [nodestyle, shift={(1.6, -3)}] {};		
		\node (5) [shift={(1.6, -2)}] {$\cdots$};
		\path [-stealth, thick]
			(2) edge [] node[left,near end] {$\pmatrix=p_1$\ \ } (3);
		\path [-stealth, thick]
			(2) edge [] node[left,near end] {$\pmatrix=p_n$\ } (4);
		\path[-stealth, thick]		
			(1) edge [loop right] node {$\pmatrix = 1, \rfunc_u = u_\init, \rfunc_l = l_\init$} (1);
		\path[-stealth, thick]		
			(3) edge [loop right] node {$\pmatrix = 1, \rfunc_u = u_1, \rfunc_l = l_1$} (3);
		\path[-stealth, thick]		
			(4) edge [loop right] node {$\pmatrix = 1, \rfunc_u = u_n, \rfunc_l = l_n$} (4);
					
		\draw [arrow] (foo)--(1);
	\end{tikzpicture}
	\caption{Computation of long-run average values.}
	\end{center}
	\vspace{-20pt}
\end{wrapfigure}
In the simpler case, the remaining model consists of the initial state $\init$ with a self-loop with probability one and $\rfunc_u = u_\init$, $\rfunc_l = l_\init$.
In this case, we have $\proplra(\pmc,\rfunc_u,\rfunc_l) = \frac{u_\init}{l_\init}$.
In the other case, the remaining model consists of the initial state $\init$ which has a probability of $p_i$ to move to one of the other $n$ remaining states $\state_i$ , $i = 1, \ldots, n$, which all have a self-loop with probability one and $\rfunc_u(\state_i) = u_i$, $\rfunc_l(\state_i) = l_i$.
In this case, we have $\proplra(\pmc,\rfunc_u,\rfunc_l) = \sum_{i=1,\ldots,n} p_i \frac{u_i}{l_i}$.

%% file: experiments.tex
\section{Experiments}
\label{sec:experiments}
We now consider four case studies that illustrate the efficiency and scalability of our approach.
Three models~\cite{helmink1993proof,ibe1990stochastic,reiter1998crowds} are taken from the
PRISM benchmark suite\footnote{\url{http://www.prismmodelchecker.org/benchmarks/}}, and the last
is taken from the authors' work on synchronisation protocols~\cite{gainer2017investigating,gainer2017power}.
All experiments were conducted on a PC with an Intel
Core i7-2600 (tm) processor at 3.4GHz, equipped with 16GB of RAM, and running Ubuntu 16.04. For
each case study we compare the performance times obtained for model analysis when using the
parametric engine of the model checker ePMC~\cite{hahn2014iscas}\footnote{\url{http://iscasmc.ios.ac.cn/?p=1241}, \url{https://github.com/liyi-david/ePMC}}, using either polynomial
fractions or DAGs to represent the functions corresponding to transition probabilities and
state rewards.
Basically, the DAG is implemented as an array of 64-bit integers.
Functions are represented as indices to this array.
4 bits describe the type of the node.
For terminal nodes, the remaining bits denote the parameter or number used.
For non-terminal nodes, $2\times 30$ bits are used to refer to the operands within the DAG.
We also compare our results to those obtained using the parametric
engine of PRISM~\cite{kwiatkowska2011prism}, and the parametric and sampling engines of Storm~\cite{dehnert2017storm}\footnote{\url{http://www.stormchecker.org/}}.

Given a parametric model, and a set of valuations for its parameters, we are interested in the
total time taken to check some property of interest for every valuation for the parameters.
Since our primary concern is the efficiency of multiple evaluations of an existing model,
we omit model construction times and restrict our analysis to the total time taken for
the evaluation of all parameter valuations. For the parametric engines of ePMC, PRISM, and Storm, we
record the total time taken for both state elimination and the evaluation of the resulting
function for all parameter valuations. For the sampling engine of Storm, we record the total
time taken for value iteration, using default settings to determine convergence.
For Storm, we set the precision to $10^{-10}$ rather than the default of $10^{-6}$.
This had a very minor influence on the runtime, and allowed a better comparison to ePMC, the results of which have a precision of $<10^{-13}$.
%\pgcomment{Need to mention that we used a precision of 1e-10 instead of the default 1e-06 for Storm.}{}

%1. num-new   ( what is this? )
%number of new transitions introduced/removed (can be negative) after elimination 
%2.    min-prod-pred-times  ( this is  |pre| * |post|, right??)
%3.    quick-target ( what is this? )
%don't know yet :)
%4.    from-target  ( what is this? )
%bfs from set of target states
%5.    random   (self-explanatory, I guess!)
%6.    node-numbers-ascending  and  node-numbers-descending  (are these simply the order in which nodes where discovered during exploration, and the reverse of that order? ) 
%bfs 

\subsection{Crowds Protocol}

The Crowds protocol~\cite{reiter1998crowds} provides anonymity for a crowd consisting of $N$
Internet users, of whom $M$ are dishonest, by hiding their communication via random routing,
where there are $R$ different path reformulates. The model is a PMC parametrised by $B = \frac{M}{M+N}$,
the probability that a member of the crowd is untrustworthy, and $P$, the probability that a member sends a
package to a randomly selected receiver. With probability $1-P$ the packet is directly delivered
to the receiver. The property of interest is the probability that the untrustworthy members
observe the sender more than they observe others.

\begin{figure}[tb]
	\centering
	\begin{tabular}{cc}
	\begin{tikzpicture}[scale=0.7]
	    \begin{axis}[grid=both,xlabel={$P$},ylabel={$B$}]
	        \addplot3[mesh, black, mesh/rows=19, mesh/cols=19, opacity=0.5] table{experiments/prism-benchmarks/crowds/surf-crowds.dat};
    	\end{axis}
	\end{tikzpicture} &
	\begin{tikzpicture}[scale=0.7]
	    \begin{axis}[grid=both,xlabel={$pK$},ylabel={$pL$}]
	        \addplot3[mesh, black, mesh/rows=19, mesh/cols=19, opacity=0.5] table{experiments/prism-benchmarks/brp/surf-brp.dat};	        
	    \end{axis}
	\end{tikzpicture} \\
	\end{tabular}
        \vspace{-7mm}
	\caption{Upper crowds protocol (L). Bounded retransmission protocol (R).}
	\label{fig:crowdsandbrp}
	% UC: (MaxGood=20,CrowdSize=5,TotalRuns=7).
	% BRP: (N=256,MAX=4)}
\end{figure}
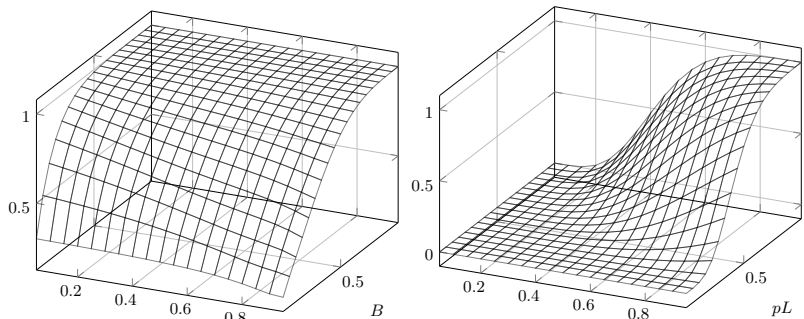

Table~\ref{tab:crowds} shows the performance
statistics for different values of $N$ and $R$, where each entry shows the total time taken
to check all pairwise combinations of values for $B, P$ taken from $0.002, 0.004, \ldots, 0.998$.
There is a substantial increase in the performance of ePMC when using non-simplified DAGs (ePMC(D)), and
using DAGs (ePMC(DS)) simplified by evaluating random points (cf.~Section~\ref{sec:dag}), instead of polynomial fractions (ePMC) to represent
functions.
Here, ePMC clearly outperforms the parametric engines of both PRISM and Storm.
In some instances, ePMC turns out to be the fastest choice, while the sampling engine of Storm proves to be faster for other instances.
Processes that exceeded the time limit of one hour are indicated by \ot, and processes
that ran out of memory are indicated by \om. In Fig.~\ref{fig:crowdsandbrp} (left) we plot
the results for $N=5$ and $R=7$.
% ... \textbf{Need to say something here...}

\begin{table}[tb]
	\setlength{\tabcolsep}{2.0pt}
	\centering
	\begin{tabularx}{\textwidth}{rrrrrrrrrr}
		\toprule
		$N$ & $R$ & States & Trans. & PRISM & ePMC & ePMC(D) & ePMC(DS) & Storm(P) & Storm(S) \\
		\midrule
		5 & 3 & 1198 & 2038 & 722 & 737 & 13 & 13 & 681 & 26 \\
		5 & 5 & 8653 & 14953 & 745 & 806 & 15 & 15 & 723 & 64 \\
		5 & 7 & 37291 & 65011 & 818 & 900 & 19 & 17 & 735 & 153 \\
		10 & 3 & 6563 & 15143 & 732 & 771 & 15 & 14 & 690 & 26 \\
		10 & 5 & 111294 & 261444 & 1146 & 910 & 23 & 16 & 712 & 63 \\
		10 & 7 & 990601 & 2351961 & \ot & \ot & 103 & 42 & 737 & 159 \\
		15 & 3 & 19228 & 55948 & 761 & 825 & 16 & 16 & 703 & 26 \\
		15 & 5 & 592060 & 1754860 & \ot & \om & 42 & 28 & 709 & 64 \\
		15 & 7 & 8968096 & 26875216 & \om & \om & \om & \om & 777 & 174 \\
		20 & 3 & 42318 & 148578 & 814 & 805 & 15 & 14 & 709 & 26 \\
		20 & 5 & 2061951 & 7374951 & \om & \om & 108 & 90 & 720 & 67 \\
%		20 & 7 &  &  &  &  &  & mem & mem &  & mem \\

		\bottomrule	
	\end{tabularx} \\
	\caption{Performance statistics for crowds protocol.}
	\label{tab:crowds}
\end{table}

\subsection{Bounded Retransmission Protocol}

The bounded retransmission protocol~\cite{helmink1993proof} divides a file, which is to be transmitted,
into $N$ chunks.  For each chunk, there are at most $\mathit{MAX}$ retransmissions over two lossy
channels $K$ and $L$ that send data and acknowledgements, respectively. 
%\pgcomment{It states that it is a PMDP in the paper. How was the non-determinism resolved to
%get the dtmc that was used in the JUnit model?}{}
The model is a PMC parametrised by $pK$ and $pL$, the reliability of the channels. We are interested in
the probability that the sender reports an unsuccessful transmission after more than 8 chunks have been sent successfully.

The performance statistics for different values of $N$ and $\mathit{MAX}$ are shown in
Table~\ref{tab:brp}, where each entry shows the total time taken
to check all pairwise combinations of values for $pK, pL$ taken from $0.002, 0.004, \ldots, 0.998$.
Here, ePMC with DAGs again has the best performance: the running time remains approximately constant when using this data structure, even for much larger problem instances.
In contrast, the running time for both engines of Storm scale linearly. Both the parametric engines of PRISM and ePMC, with
polynomial fraction representation, run out of memory for all larger problem instances.

Figure~\ref{fig:crowdsandbrp} (right) plots the results obtained for
$N=256$ and $MAX=4$.
As we see the probability of interest first increases with increasing channel reliability, but then decreases again.
The reason is that, on the one hand, if the channel reliability is low, then we do not send many chunks successfully.
On the other hand, if the channel reliability is high, then it is unlikely that the transmission will fail in the end.

\begin{table}[tb]
	\setlength{\tabcolsep}{2.0pt}
	\centering
	\begin{tabularx}{\textwidth}{rrrrrrrrrr}
		\toprule
		$N$ & $\mathit{MAX}$ & States & Trans. & PRISM & ePMC & ePMC(D) & ePMC(DS) & Storm(P) & Storm(S) \\
		\midrule
		64 & 4 & 4139 & 5543 & 1029 & 1016 & 36 & 38 & 991 & 160 \\
		64 & 5 & 4972 & 6695 & 1145 & 1118 & 36 & 33 & 1021 & 188 \\
		256 & 4 & 16427 & 22055 & \om & \om & 48 & 40 & 3332 & 403 \\
		256 & 5 & 19756 & 26663 & \om & \om & 35 & 15 & \ot & 318 \\
		512 & 4 & 32811 & 44071 & \om & \om & 29 & 19 & \ot & 491 \\
		512 & 5 & 39468 & 53287 & \om & \om & 28 & 23 & \ot & 596 \\
		\bottomrule	
	\end{tabularx} \\
	\caption{Performance statistics for bounded retransmission protocol.}
	\label{tab:brp}	
\end{table}

\subsection{Cyclic Polling Server}
This cyclic server polling model~\cite{ibe1990stochastic} is a model of a network, described as a continuous-time Markov chain.
There are two parameters, $\mu$ and $\gamma$.
The model consists of one server and $N$ clients.
When a client is idle, then a new job arrives at this client with a rate of $\mu/N$.
The server `polls' the clients in a cyclic manner.
At each point of time, it observes a single client.
If there is a job waiting for a given client, the server servers its job (provided there is one) with a rate of $\mu$.
When the client it observes is idle, then the server moves on to observe the next client with a rate of $\gamma$.
Even though our method targets discrete-time models, we can handle this model by computing the embedded DTMC.

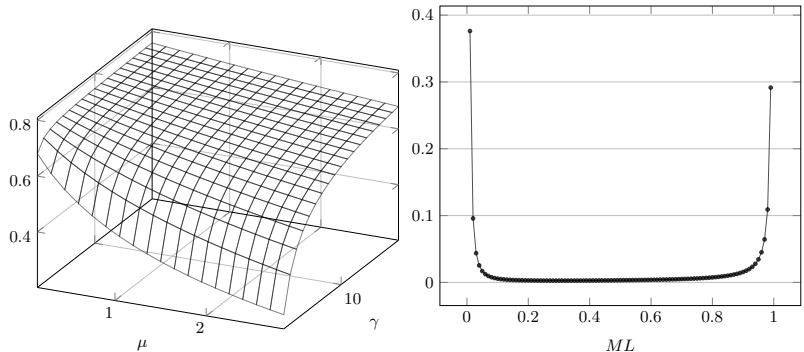
\begin{figure}[tb]
	\centering
	\begin{tabular}{cc}
	\begin{tikzpicture}[scale=0.7]
	    \begin{axis}[grid=both,xlabel={$\mu$},ylabel={$\gamma$}]
	        \addplot3[mesh, black, mesh/rows=19, mesh/cols=19, opacity=0.5] table{experiments/prism-benchmarks/polling/poll8/surf-poll8.dat};
    	\end{axis}
	\end{tikzpicture} &
	\begin{tikzpicture}[scale=0.7]
	    \begin{axis}[xmajorgrids=false,ymajorgrids=true,xlabel={$ML$},ylabel={}]
	        \addplot[mark size=1.0pt, mark=*, opacity=0.7] table {experiments/synch/synch-2.dat};	        
    	\end{axis}
	\end{tikzpicture} \\		
	\end{tabular}
        \vspace{-5mm}
	\caption{Cyclic polling server (L). Synchronisation model (R).}
	%\caption{Cyclic polling server, $N=8$ (L). Synchronisation model (R).}	
	\label{fig:cyclicandsynch}	
	% SM: (N=6,T=8)).
	% CP  (N=4): 
\end{figure}

In this case study, we consider the probability that, in the long run, Station 1 is idle.  That is, the expected limit average of the time that Station~1, or, due to symmetry, any other station, is idle.
We compute this long-run average value using the method described in Section~\ref{sec:lra}.
Probabilities are displayed as a function of the parameters in Figure~\ref{fig:cyclicandsynch}, and Table~\ref{tab:cyclic} shows how the various tools perform on this benchmark.
% With increasing $\mu$, the probability that Station 1 is idle decreases.
% This is because the arrival rate of arrival of jobs (for idle clients) is $\mu/N$.
% Even though the server processes jobs with a rate of $\mu$, this cannot fully compensate for the increasing rate of incoming jobs.
With increasing $\gamma$ the likelihood that Station 1 is idle increases:
if we increase $\gamma$, then the server will more quickly find stations to be served.
As the long-run average idle time only depends on the rate between $\mu$ and $\gamma$, the likelihood that Station 1 is idle falls with increasing $\mu$.

For the current configuration, classic parametric model checking does not seem to be advantageous.
Using our DAG-based implementation, however, is much more efficient than classic parametric model checking, but it is space consuming.
With the chosen number of parameter instantiations, our method does not quite compete with non-parametric model checking.

\begin{table}[tb]
	\setlength{\tabcolsep}{3.4pt}
	\centering
	\begin{tabularx}{\textwidth}{rrrrrrrrr}
		\toprule
		$N$ & States & Trans. & PRISM & ePMC & ePMC(D) & ePMC(DS) & Storm(P) & Storm(S) \\
		\midrule
%		4 & 96 & 272 & 0 & 421 & 136 & 3 & 3 \\
%		5 & 240 & 800 & 1 & \ot & \ot & 3 & 3 \\		
%		6 & 576 & 2208 & 1 & \ot & \ot & 3 & 3 \\
%		7 & 1344 & 5824 & 2 & 2126 & \ot & 8 & 7 \\
%		8 & 3072 & 14848 & 4 & 1542 & \ot & 34 & 43 \\		
%		9 & 6912 & 36864 & 10 & 2487 & \ot & \om & \om \\		
%		4 & 96 & 272 & 0 & 416 & 124 & 2 & 2 & 893 \\
%		5 & 240 & 800 & 1 & \ot & \ot & 2 & 2 & \ot \\
%		6 & 576 & 2208 & 1 & 3457 & \ot & 3 & 3 & \ot \\
%		7 & 1344 & 5824 & 2 & \ot & \ot & 9 & 7 & \ot \\
%		8 & 3072 & 14848 & 4 & 986 & \ot & 20 & 22 & \ot \\
%		9 & 6912 & 36864 & 8 & 1498 & \ot & \om & \om & \ot \\
%		4 & 96 & 272 &  & 535.524(116.673) &  240.602(118.59) & 21.177000000000000001(18.723) & 199.191(196.598) & 291.14900000000000002(108.515) & 7.9920000000000000003(7.971) \\
		4 & 96 & 272 & 1166 & 888 & 14 & 14 & 953 & 50 \\
		5 & 240 & 800 & \ot & \ot & 28 & 25 & \ot & 121 \\
		6 & 576 & 2208 & 3550 & \ot & 108 & 102 & \ot & 305 \\
		7 & 1344 & 5824 & 1399 & \ot & 759 & 736 & \ot & 801 \\
		8 & 3072 & 14848 & 1052 & \ot & \ot & \ot & \ot & 1991 \\
		9 & 6912 & 36864 & \ot & \ot & \om & \om & \ot & \ot   \\
		\bottomrule	
	\end{tabularx} \\
	\caption{Performance statistics for cyclic polling server.}
	\label{tab:cyclic}	
\end{table}

\subsection{Oscillator Synchronisation}

The models of~\cite{gainer2017investigating,gainer2017power} encode the behaviour of
a population of $N$ coupled nodes in a network. Each node has a clock that
progresses, cyclically, through a range of discrete values $1,\ldots,T$. At the
end of each clock cycle a node transmits a message to other nodes in the network. Nodes that
receive this message adjust their clocks to more closely match those of the firing node.
The model is a PMC, parametrised by the likelihood $\textit{ML}$ that a firing
message is lost in the communication medium. The property of interest is the expected
power consumption of the network (in Watt-hours) to reach a state, where the clocks of
all nodes are synchronised.

Table~\ref{tab:synch} shows the results for different values of $N$ and $T$,
where each entry shows the total time taken to check all values of $\textit{ML}$
%taken from $0.00001, 0.00002, \ldots, 0.99999$.
taken from $10^{-5}, 2 \cdot 10^{-5}, \ldots, 1 - 10^{-5}$.
% \textbf{Something about table of results here, when they are finished!}

Figure~\ref{fig:cyclicandsynch} (right) plots the results obtained for
$N=6$ and $T=8$.
For extremal values of $\textit{ML}$, the network
is expected to use much more energy to synchronise, because the expected time required
for this to occur increases. Very high values of
$\textit{ML}$ result in nearly all firing messages being lost, and hence nodes
cannot communicate well enough to coordinate, while very low values of $\textit{ML}$
lead to perpetually asynchronous states for the network, an artefact of the
discreteness of the clock values~\cite{gainer2017investigating}.

\begin{table}[tb]
	\setlength{\tabcolsep}{2.8pt}
	\centering
	\begin{tabularx}{\textwidth}{rrrrrrrrrr}
		\toprule
		$N$ & $T$ & States & Trans. & PRISM & ePMC & ePMC(D) & ePMC(DS) & Storm(P) & Storm(S) \\
		\midrule
4 & 6 & 218 & 508 & 280 & 304 & 6 & 4 & 6 & 20 \\
4 & 7 & 351 & 822 & 302 & 399 & 7 & 4 & 6 & 28  \\
4 & 8 & 535 & 1257 & 542 & 1520 & 9 & 5 & 13 & 37 \\
5 & 6 & 449 & 1179 & 354 & 499 & 10 & 5 & 11 & 39 \\
5 & 7 & 799 & 2094 & 1694 & \ot & 11 & 6 & 34 & 60 \\
5 & 8 & 1333 & 3533 & \ot & \ot & 17 & 8 & 137 & 90 \\
6 & 6 & 841 & 2491 & 1070 & \ot & 12 & 7 & 48 & 74 \\
6 & 7 & 1639 & 4820 & \ot & \ot & 19 & 8 & 239 & 130 \\
6 & 8 & 2971 & 8871 & \ot & \ot & 33 & 10 & 2311 & 211 \\
		\bottomrule	
	\end{tabularx} \\
	\caption{Performance statistics for synchronisation model.}
	\label{tab:synch}	
\end{table}

In this case study, the DAG-based method, in particular with random points evaluation, performs best, followed by the sampling-based method of Storm.
We note that the time required for each value iteration is relatively high, while the cost of evaluating a point for the DAG-based method is quite low.
Therefore, the advantage of our approach would have been even more pronounced, if we had  evaluated more instantiations in the experiments above.
The method of choice thus depends mostly on whether such a high number of instantiations is required.

We have performed value iteration with a (local) precision of $10^{-10}$ for Storm.
This does, however, not guarantee any global precision~\cite{BKLPW17}.
Obtaining guaranteed results using value iteration is relatively expensive while, as discussed in Section~\ref{sec:dag},
extending our approach to obtain conservative guarantees is relatively simple---and inexpensive---to achieve by using basic interval arithmetic.

%\textbf{Need to discuss number of points - increase in time will be linear for value iteration, but relatively free for DAGs}

%\textbf{Need to discuss quality of results - we use PRISM default (unsafe) value iteration, with safe value iteration we need ~4x the time CITE}

%\textbf{We are only limited by machine precision for our results....}
%\textbf{Easy to extend to using interval arithmetic for evaluation, which should not increase overheads}
%Moritz: removed, I added a discussion to DAG section instead

\subsection{Heuristics}
\label{sec:heuristics}

\input{heuristics}

In Table~\ref{tab:heuristics}, we compare the different heuristics described.
We have applied each of them for each considered model, and provide the time in seconds required for medium-sized instances.
As seen, it turns out that \ebfstarg\ is in general a good choice.
In one case, however, \enumnew\ turns out to be faster.

\begin{table}[tb]
	\setlength{\tabcolsep}{2.5pt}
	\centering
	\begin{tabularx}{\textwidth}{lrrrrrrrrr}
		\toprule
		Model & & \multicolumn{6}{c}{Elimination Heuristic} \\ 		\cline{3-8}
		& & \enumnew & \eminprod & \ebfstarg & \erandom & \ebfs & \ereversebfs \\		
		\midrule
		Crowds & ($N{=}10$,$R{=}5$) & 19 & 103 & 5 & 14 & 22 & 6 \\
		BRP & ($N{=}512$,$\textit{MAX}{=}5$) & 4 & 11 & 4 & \om & 5 & 4 \\		
		Cyclic & ($N{=}7$) & 7 & 9 & 8 & 8 & 8 & 8\\
		Synch & ($N{=}6,T{=}8$) & 18 & 18 & 17 & 19 & 18 & 17 \\		
		\bottomrule	
	\end{tabularx} 
	\caption{Performance statistics for different heuristics.}
	\label{tab:heuristics}
\end{table}

%% file: heuristics.tex
An important consideration when performing state elimination is the order, in which
different states are eliminated from the graph. Using different elimination orders
to evaluate the same model can result in functions, whose representations
(nominator-denominator or DAGs) vary greatly in size, and hence also in the
corresponding memory footprint and analysis time. Heuristics for efficient
state elimination have been studied in automata theory, to obtain shorter regular
expressions from finite-state automata~\cite{han2007obtaining,han2013state}, and
in graph theory, for efficient peeling of a probabilistic network~\cite{harbron1995heuristic}.
We employ the following heuristics, consisting of both existing schemes taken from the
literature, and novel schemes that prove to be effective for some models. 
\begin{itemize}
\item \enumnew:
each state is weighted by the number of new transitions that are introduced
to the model when that state is eliminated. That is, we consider each
predecessor-successor pair for that state, and add one to the weight if
the transition from the predecessor to the successor was not already defined
in the underlying graph before state elimination. States with the lowest
weight are eliminated first. The aim here is to minimise the total number of
transitions as elimination progresses. 
\item \eminprod:
similarly to \enumnew, we consider each predecessor-successor pair.
However, one is added to the weight irrespective of whether that transition already
existed in the underlying graph. Again states with the lowest weight are
considered first.
\item \ebfstarg:
states are eliminated in the order in which they are discovered when
conducting a breadth-first search backwards from the target states.% of interest.
%\textbf{TODO motivation?}
\item \erandom:
a state is selected uniformly at random for elimination from the set of
remaining states.
\item \ebfs:
states are eliminated in the order in which they are discovered when
conducting a breadth-first search from the initial state(s) of the model.
%\textbf{TODO motivation?}
\item \ereversebfs:
similar to \ebfs, except states are eliminated in reverse order.
\end{itemize}
% For each case study we indicate the applied heuristic, and at the end of this
% section we compare the effectiveness of using different heuristics for the 
% analysis of sample instances of each of the four models under consideration. 
% %\textbf{TODO motivation?}
% 
% \textbf{TODO Need to state which elimination order we used for each case study}

%% file: conclusion.tex
\section{Conclusion and Future Work}
\label{sec:conclude}

We have implemented an approach for the evaluation of parametric Markov chains that exploits the synergies of using DAGs in a state-elimination based analysis and using DAGs in an encoding of rational functions as arithmetic circuits.
Our experimental evaluation suggests that these two approaches integrated so seamlessly that they often provide a notable speedup.
The nicest observation is that this seems so natural in hindsight that it is almost more surprising that this has not been attempted before than that it works so well.
We therefore hope to have discovered one of these simple and natural approaches that will stand the test of time.

The next step in exploiting our approach could be an integration into applications.
One of the applications we have in mind is to use it in the context of parameter extraction, which we expect to work similar to Model extraction, for online Model checking.
The growing knowledge of the model can be used to refine or adjust the parameters in this application.
Our application can help to provide the speed required to make the approach scale, and to keep the analysis and, if required, the visualisation%
\footnote{We obtain the probablities and expected rewards as functions of the parameters. This can be visualised, just as we have visualised this in Section \ref{sec:experiments}.}
of the effect of the learnt parameters (and the confidence area around them) efficient.

We also note that interval arithmetic could be used to evaluate \emph{boxes}---hyperrectangles $[a_1,b_1] \times \cdots [a_n,b_n]$
of parameter ranges---so as to obtain bounds on the lower and upper values taken by any occurring function value in the box.
This approach could be used instead of using SMT solvers (as in~\cite{DehnertJ0CVKAB16,HahnHZ11}) to decide PCTL properties.
A similar approach to avoid using SMT solvers has been proposed~\cite{QuatmannD0JK16}, which is however not based on computing a function depending on the parameters but on value iteration.
We assume that the DAG-based approach will perform better when a high coverage of the parameter space is required.

It would also be straightforward to parallelise evaluation of points using SIMD approaches such as GPGPU.

%% file: acknowledgements.tex
\section{Acknowledgements}
\vspace{-3mm}
This work was supported by the Sir Joseph Rotblat Alumni Scholarship at Liverpool,
EPSRC grants EP/M027287/1 and EP/N007565/1, and by the Marie Sk{\l}odowska Curie Fellowship \emph{Parametrised Verification and Control}.